\documentclass[12pt]{article}
\usepackage{epsfig,amsmath,amssymb,stmaryrd,enumerate}

\advance \textheight by \topskip
\advance \textheight by 1pt
\makeatother

\newtheorem{lemma}{Lemma}
\newtheorem{theorem}{Theorem}
\newtheorem{corollary}{Corollary}
\def\bea{\begin{eqnarray}}
\def\ena{\end{eqnarray}}
\def\non{\nonumber}
\newcommand{\bc}[2]{
\left(
\begin{array}{c}{#1}\\{#2}\end{array}
\right)}

\title{
Discretization of  Baker--Akhiezer Modules and Commuting Difference Operators in Several Discrete Variables
}

\author{
Andrey Mironov\thanks{e-mail:mironov@math.nsc.ru}\\
Sobolev Institute of Mathematics,\\
630090 Novosibirsk, Russia\\
 and \\
Laboratory of Geometric Methods in \\
Mathematical Physics,\\
 Moscow State University
\\
\\
Atsushi Nakayashiki\thanks{
e-mail: atsushi@tsuda.ac.jp}\\
Department of Mathematics,\\
Tsuda College\\
Kodaira, Tokyo, Japan\\
}
\date{}

\begin{document}

\maketitle

\centerline{
Dedicated to Viktor
Matveevich Buchstaber on his seventieth birthday
}

\vskip10mm

\begin{abstract}
We introduce the notion of discrete Baker-Akhiezer (DBA) modules, which are modules over the ring of difference operators, as a certain
discretization of Baker-Akhiezer modules which are modules over the ring of differential operators. We use it to construct commuting difference
operators with matrix coefficients in several discrete variables.
\end{abstract}
\clearpage

\section{Introduction}

In this paper we introduce the notion of discrete Baker--Akhiezer 
modules and, with the help of it, construct commutative rings of 
 difference operators with matrix coefficients in several discrete variables from certain algebraic varieties.

We firstly recall some basic facts on commuting difference operators in one variable.
Common eigen functions of two commuting difference operators

\begin{equation}\label{e1}
 L_1=\sum^{N_+}_{i=N_-}v_i(n)T^i,\ \ L_2=\sum^{M_+}_{i=M_-}u_i(n)T^i
\end{equation}
are parametrized  by points of some algebraic curve $\Gamma$
$$
 L_1\psi(n,P)=\lambda(P)\psi(n,P),\ L_2\psi(n,P)=\mu(P)\psi(n,P).
$$
Krichever and Novikov \cite{KrN} proved that on $\Gamma$ there are
points $P_1,\dots,P_k$  such that the whole commutative ring of
difference operators, containing $L_1$ and $L_2$, is isomorphic
to the ring of meromorphic functions with poles only at 
$P_1,\dots,P_k$. In the case $k=2$ (two-point construction) 
explicit forms of operators were found in \cite{Kr,Mum}. 
The theory of $n$-point operators was developed in \cite{KrN}. 
Krichever and Novikov classified one-point operators of rank $l$ and found
operators of rank two corresponding to the spectral curve of genus one.
The theory of such operators is connected with the theory of higher
rank algebro-geometric solutions of 2D-Toda chain \cite{KrN}. In
\cite{Mir}  Krichever--Novikov operators with polynomial coefficients
are found.

In the case of operators, either differential or difference, of several
variables, there is no classification theorem up to now 
(some results in this direction see in \cite{Kr1,Zh,KOZ}). 
The main difficulty is as follows. 
If ordinary difference operators (\ref{e1}) have a family of
common eigen functions parametrized by an algebraic curve with $\lambda$ and $\mu$ being functions on it, then they commute. 
On the other hand, in the case of operators of 
several variables, only the existence of a big family of common eigen
functions is not enough for commutativity.
 For example, it is not
difficult to construct operators possessing a family of common eigen
functions parametrized by points of an algebraic variety which do not commute. 
This is a major difference between one
 and higher dimensional cases.

Then the main question is how many common eigen functions are enough for the commutativity in the multi-dimensional case.
An answer to this question is partially given
in the papers of the second author \cite{N1,N2}. 
In these papers the notion of Baker--Akhiezer (BA) modules over the ring of differential operators are introduced.
It allows one to obtain commuting differential operators in several variables with matrix coefficients.

In this paper we introduce a discrete analogue of BA modules.
It makes it possible to construct commuting partial difference operators 
with matrix coefficients as an analogue of the construction of commuting
differential operators.

Let ${\hat M}$ be a set of functions $\psi(n,P),$ $n\in{\mathbb Z}^g$, $P\in X,$ where $X$ is an algebraic
variety (spectral variety). We assume that elements of ${\hat M}$ have the  following properties.
\vskip5mm
\noindent
{\bf 1}. $T_i\psi(n,P)\in {\hat M},$ where $T_i$ is a shift
operator on the $i$-th discrete variable of $n=(n_1,...,n_g)$.
\vskip2mm
\noindent
{\bf 2}. $f(n)\psi(n,P)\in {\hat M}$, where $f(n)$ is an arbitrary function
from a certain class.
\vskip2mm
\noindent
{\bf 3}. $\lambda(P)\psi(n,P)\in {\hat M}$, where $\lambda(P)$ is a meromorphic function on $X$ with poles only on some fixed subvariety  
$Y$ of $X$.
\vskip2mm

Let $A_Y$ be the ring of meromorphic functions on $X$ with poles only on $Y$ and ${\cal T}_g={\hat {\cal K}}[T_1,\dots,T_g]$ the ring of difference operators, where ${\hat {\cal K}}$ is a ring of certain functions on ${\mathbb Z}^g$.
The properties 1-3 imply that ${\hat M}$ is a module over ${\cal T}_g$ and, at the same time, over $A_Y$.
We call ${\hat M}$ a Discrete Baker--Akhiezer (DBA) module.

Suppose that ${\hat M}$ is a free ${\cal T}_g$-module of finite rank. Then the DBA-module allows us to construct commuting difference operators in several variables.
Indeed, let us choose a free basis $\psi_1,\dots,\psi_N$ in ${\hat M}$ and
consider the vector valued function $\Psi={}^t(\psi_1,\dots,\psi_N)$. 
Then for $\lambda\in A_Y$ there exists uniquely a difference operator with matrix coefficients $D(\lambda)$ such that
$$
 D(\lambda)\Psi=\lambda\Psi,
$$
since ${\hat M}$ is a free ${\cal T}_g$ module.
Similarly, for $\mu\in A_Y$, we have 
$$
 D(\mu)\Psi=\mu\Psi.
$$
Operators $D(\lambda)$ and $D(\mu)$ commute, since ${\hat M}$ is free and
$\lambda$, $\mu$ do not depend on the discrete variable $n$.
 It means that the family $\{\Psi(n,P)\}$ of common eigen vector valued functions parametrized by
points of $X$ is large enough and from commutativity on $\{\Psi(n,P)\}$ follows that operators commute on the whole space of vector valued functions.

We construct examples of free DBA modules of finite rank
and commuting difference operators from abelian varieties with non-singular
theta divisors and certain rational varieties as certain discretizations 
of the corresponding BA modules. We show that a basis of a BA-module gives
a basis of the corresponding DBA module. This kind of structure that 
solutions of continuous system directly give solutions of the corresponding 
discrete system is well known in soliton equations \cite{Miwa,DJM}.

The present paper is organized as follows.
In section 2 we construct DBA modules explicitly and  give main theorems. The DBA modules are formulated as certain discretizations of
 Baker-Akhiezer ${\cal D}$ modules. Proofs of theorems are given in section 3. In section 4 we give examples of explicit forms of operators.

\section{Construction of free DBA-modules}
In this section we give two examples of free DBA-modules which are constructed from  Abelian varieties and certain rational varieties.
 In the first
case elements of DBA-modules and coefficients of difference operators are expressed in terms of theta-functions and in the second
case the corresponding objects are expressed by elementary functions.
All theorems in this section can be proved using the results on their differential analogues. Proofs themselves are given in section 3.

\subsection{DBA-modules on Abelian varieties}

Let $\tau$ be a point of the Siegel upper half space, $\theta_{a,b}
(z,\tau)$ the
Riemann's theta function with the characteristic
${}^t({}^ta,{}^tb)$, $a,b\in {\mathbb R}^{g}$, 
$X={\mathbb C}/({\mathbb Z}^g+\tau{\mathbb Z}^g)$,
$\Theta\subset X$ the theta divisor specified by the zero set of $\theta(z):=\theta_{0,0}(z,\tau)$ and ${\cal L}_c$, $c\in {\mathbb C}^g$,
the flat line bundle on $X$ for which
 $\theta(z+c)/\theta(z)$ is a meromorphic section.
A meromorphic section of ${\cal L}_c$ is identified
with a meromorphic function $f(z)$ on ${\mathbb C}^g$
satisfying the condition
\bea
&&
f(z+m+\tau n)=exp(-2\pi i{}^tn c)f(z),
\label{trf-Lc}
\ena
for any $m,n\in {\mathbb Z}^g+\tau {\mathbb Z}^g$.

Let $L_c(m)$ be the space of meromorphic sections of ${\cal L}_c$ with
poles only on $\Theta$ of order at most $m$ and $L_c=\cup_{m=0}^\infty L_c(m)$. A basis of $L_c(m)$ is given quite explicitly.
Namely, for a nonnegative integer $m$ and $a\in {\mathbb Z}^g/m{\mathbb Z}^g$
we set
\bea
&&
F_{m,a}(z,c)=\theta_{a/m,0}(mz+c,m\tau)/\theta(z)^m.
\non
\ena
Then the set of functions $\{F_{m,a}(z,c)\}$ is a basis of $L_{c}(m)$.

We denote by ${\cal K}$
the ring of meromorphic functions on ${\mathbb C}^g$.
We denote the variable of a function of ${\cal K}$ by
$x=(x_1,...,x_g)$.
Define the space $M_c$ by
\bea
&&
M_c=\cup_{m=0}^\infty M_c(m),\quad
M_c(m)=\sum_a {\cal K}F_{m,a}(z,c+x).
\non
\ena
This is nothing but the underlying space of the Baker-Akhiezer module
of $(X,\Theta)$ \cite{N1}. We shall discretize it as follows.

For a function $F(z,x)$ define the operator $T_i$ by
\bea
&&
T_iF(z,x)=F(z,x+h_ie_i)\frac{\theta(z-h_ie_i)}{\theta(z)},
\quad
F(z,x)\in M_c,
\non
\ena
where $e_i$ is the $i$-th unit vector of ${\mathbb C}^g$  and
$h_i\in {\mathbb C}$ is a parameter. It is easy to see that
$T_i$ acts on $M_c$, since it preserves the relation (\ref{trf-Lc})
for $L_{c+x}$.

For $f(x)\in{\cal K}$ we associate the map
${\hat f}:{\mathbb Z}^g\rightarrow {\cal K}$ by
\bea
&&
{\hat f}(n)=f(x+nh),
\non
\ena
where $n=(n_1,...,n_g)$ and $nh=(n_1h_1,...,n_gh_g)$.
We identify the map ${\hat f}$ with its value ${\hat f}(n)$.
Let
\bea
&&
{\hat {\cal K}}=\{{\hat f}(n)\vert f\in {\cal K}\}.
\non
\ena
The space ${\cal K}$ naturally becomes a ring which we consider
the ring of discrete functions with the discrete variable
$n\in {\mathbb Z}^g$.

For a non-negative integer $m$ and $a\in {\mathbb Z}^g/m{\mathbb Z}^g$
we define the map ${\hat F}_{m,a}:{\mathbb Z}^g\rightarrow M_c$ by
\bea
&&
{\hat F}_{m,a}(n)=T^nF_{m,a}(z,c+x),
\non
\ena
where $T^n=T_1^{n_1}\cdots T_g^{n_g}$.
We identify the map ${\hat F}_{m,a}$ and its value ${\hat F}_{m,a}(n)$.
We write ${\hat F}_{m,a}(n,z)$ if it is necessary to indicate the dependence on the variable $z$.

Now we define the discrete Baker-Akhiezer module ${\hat M}_c$ by
\bea
&&
{\hat M}_c=\cup_{m=0}^\infty {\hat M}_c(m),
\quad\quad
{\hat M}_c(m)=\sum_{a\in {\mathbb Z}^g/m{\mathbb Z}^g}
{\hat {\cal K}}{\hat F}_{m,a}(n).
\non
\ena
Explicitly
\bea
&&
{\hat M}_c(m)=\sum_a {\hat {\cal K}}
\frac{\theta_{a/m,0}(mz+c+x+nh,m\tau)}{\theta(z)^m}
\prod_{j=1}^g
\left(\frac{\theta(z-he_j)}{\theta(z)}\right)^{n_j}.
\label{explicit-Mchm}
\ena

We give an example of the elements in
${\hat M}_c(m)$ with $m=1,2$.
\vskip2mm
\noindent{\bf Example.}
$$
 \frac{\theta(z+c+x+nh)}{\theta(z)}\prod_{j=1}^g
\left(\frac{\theta(z-he_j)}{\theta(z)}\right)^{n_j}
 \in {\hat M}_c(1),
$$
$$
 \frac{\theta(z+c+x+nh+\beta)\theta(z-\beta)}{\theta^2(z)}
\prod_{j=1}^g
\left(\frac{\theta(z-he_j)}{\theta(z)}\right)^{n_j}\in
{\hat  M}_c(2),
$$
where $\beta$ is an arbitrary constant from ${\mathbb C}^g$. The first example corresponds to $m=1$, $a=0$ in (\ref{explicit-Mchm}).
\vskip2mm

The operator $T_i$ acts on ${\hat M}_c$ as the shift operator:
\bea
&&
T_i\left({\hat f}(n){\hat F}_{m,a}(n)\right)
={\hat f}(n+e_i){\hat F}_{m,a}(n+e_i).
\non
\ena

Let ${\cal T}_g={\hat{\cal K}}[T_1,...,T_g]$ be the ring of difference
operators with the coefficients in ${\hat{\cal K}}$. Then ${\hat M}_c$ becomes a ${\cal T}_g$ module.

Let $A=L_0$ be the ring of meromorphic functions on $X$ which is regular on $X\backslash \Theta$. Obviously the space $L_{c+x}$ is an $A$ module. It follows that the ring $A$ also acts on
${\hat M}_c$.
In fact, for $f(z)\in A$, we have
\bea
&&
f(z)F_{m,a}(z,c+x)=\sum_{m',a'}f_{m',a'}(x)F_{m',a'}(z,c+x),
\label{action}
\ena
for some $f_{m',a'}(x)\in {\cal K}$, since $L_{c+x}$ is an $A$-module.  Notice that the multiplication by $f(z)$ commutes with the action of $T_i$. Therefore, applying $T^n$ to (\ref{action}),
we have
\bea
&&
f(z){\hat F}_{m,a}(n)=\sum_{m',a'}{\hat f}_{m',a'}(n){\hat F}_{m',a'}(n),
\non
\ena
which shows $f(z){\hat M}_c\subset {\hat M}_c$.
Consequently ${\hat M}_c$ is a $({\cal T}_g,A)$ bi-module.

In the following we assume that $\Theta$ is non-singular.
Then our first theorem is

\begin{theorem}\label{th-1}
 For an uncountable number of $h\in ({\mathbb C}^{\ast })^g$ the module ${\hat M}_c$ is a free ${\cal T}_g$-module of rank $g!$, where ${\mathbb C}^\ast
={\mathbb C}\backslash\{0\}$.
\end{theorem}

\begin{corollary}
For values of $h$ specified in Theorem \ref{th-1} 
there exists a ring mono-morphism
$$
 A\rightarrow{\rm Mat}(g!,{\cal T}_g).
$$
\end{corollary}

\subsection{DBA-modules on rational varieties}

We construct a rational spectral variety $\Gamma$ from
${\mathbb C}P^{1}\times{\mathbb C}P^{g-1}$ by
identifying two hypersurfaces.
In general we denote a point of the projective space
${\mathbb C}P^{m-1}$ by $[t_1,...,t_m]$ while a point
of the $m$ dimensional affine space by $(t_1,...,t_m)$.

Let us fix  $a_1$, $a_2$, $b_1$, $b_2 \in {\mathbb C}$ such that
$(a_{i},b_{i})\ne (0,0)$ and $[a_1,b_1]
\neq [a_2,b_2]$. Let $\mathcal{P}$ be a non-degenerate linear map
$\mathcal{P}:{\mathbb C}^{g}\rightarrow{\mathbb C}^{g}$, $\lambda_j$ and $v_j, j=1\dots,g$, the eigenvalues and the eigenvectors of $\mathcal{P}$ respectively.
We assume that $\lambda_i\ne\lambda_j$ for $i\ne j$.
 Denote the induced map ${\mathbb C}P^{g-1}\rightarrow {\mathbb C}P^{g-1}$ by the same symbol $\mathcal{P}$.

We set
$$
 \Gamma={\mathbb C}P^{1}\times{\mathbb C}P^{g-1}/\{([a_{1},b_{1}],t)\sim([a_{2},b_{2}],\mathcal{P}(t)), \
 t\in{\mathbb C}P^{g-1} \}.
$$
Then on $\Gamma$ there is a structure of an
algebraic variety \cite{MM}.

Let $f(P),f_1(P),\dots,f_g(P)$ be the functions of
$P=(z_1,z_2,t_1,...,t_g)\in {\mathbb C}^{g+2}$ of the form

\begin{equation} \label{eq2}
f(z_{1},z_{2},t_{1},\dots,t_{n})=\sum_{i=1}^{g} (\alpha_{i} z_{1}
t_{i}+\beta_{i} z_{2} t_{i}), \ \alpha_{i},\beta_{i}\in{\mathbb
C},
\end{equation}

\begin{equation} \label{eq2.5}
f_{i}(z_{1},z_{2},t_{1},\dots,t_{g})=\sum_{k=1}^{g} \left (
\alpha_{ik} z_{1} t_{k}+\beta_{ik} z_{2} t_{k} \right ),\ \alpha_{ik},\beta_{ik}\in{\mathbb C},
\end{equation}
such that
\begin{equation} \label{eq5}
f(a_{1},b_{1},v_j)\neq 0 ,\ j = 1,\dots,g,
\end{equation}
\begin{equation} \label{eq3}
f(a_{1},b_{1},t)-A f(a_{2},b_{2},\mathcal{P}(t))=0,
\end{equation}
\begin{equation} \label{eq6}
f_{i}(a_{1},b_{1},t)-c_if_{i}(a_{2},b_{2},\mathcal{P}(t))=0,
\end{equation}
for fixed $A,c_1,\dots,c_g\in{\mathbb C}^*$ and every $t=(t_{1},\dots,t_{g}).$
Moreover we choose parameters $(\alpha,\beta),(\alpha_i,\beta_i)$
 in general position,
which means that the parameters belong to some open domain (see \cite{MM} to specify this domain).

 According to (\ref{eq3}) the equation
$$
f(z_{1},z_{2},t_{1},\dots,t_{g})=0
$$
correctly defines a hypersurface in $\Gamma$.

Let us fix $\Lambda \in {\mathbb C^*}$.
The discrete Baker-Akhiezer module ${\hat M}_\Lambda$ is similarly
defined to the case of Abelian varieties as the discretization of
the Baker-Akhiezer module constructed in \cite{MM}.
It is  defined directly by
\bea
&&
{\hat M}_{\Lambda}=\cup_{k=0}^\infty {\hat M}_{\Lambda}(k),
\label{Mlh-gamma}
\\
&&
{\hat M}_{\Lambda}(k)=\left\{\psi(n,P)=
\frac{h(n,P)}{f(P)^k}\prod_{j=1}^g\left(\frac{f_j(P)}{f(P)}\right)^{n_j}\right\},
\label{Mlhk-gamma}
\ena
where $h(n,P)=h(n,z_1,z_2,t)$ is of the form

\begin{equation} \label{eq9}
h(n,P)=\sum_{0\leq j \leq k,|\alpha|=k} h_{j
\:\alpha}(n)\,z_{1}^{j}z_{2}^{k-j}\, t^{\alpha},
\end{equation}
where $\alpha=(\alpha_{1},\dots,\alpha_{g})$,
$t^{\alpha}=t_{1}^{\alpha_{1}}\!\cdot \dots \cdot
t_{g}^{\alpha_{g}}$,
and satisfies the equation

\begin{equation} \label{eq8}
\psi(n,a_{1},b_{1},t) - \Lambda \psi(n,a_{2},b_{2},\mathcal{P}(t)) =
0.
\end{equation}

Note that by (\ref{eq8}) we have
$$
 \frac{h(n,a_1,b_1,t)}{f(n,a_1,b_1,t)^k}-
\Lambda\frac{h(n,a_2,b_2,\mathcal{P}(t))}{f(n,a_2,b_2,\mathcal{P}(t))^k}
\prod_{j=1}^g
 \left(\frac{A}{c_j}\right)^{n_j}=0.
$$

According to (\ref{eq3}), (\ref{eq6}), (\ref{eq8}), if $\psi\in {\hat M}_{\Lambda}(k)$, then $T_j\psi=\psi(n+e_i,P)\in {\hat M}_{\Lambda}(k+1)$.
Consequently, we have $g$ mappings
$$
 T_j : {\hat M}_{\Lambda}(k) \rightarrow {\hat M}_{\Lambda}(k+1), \;
j=1,\dots,g.
$$

\begin{theorem}\label{th-2}
For an uncountable number of $h\in ({\mathbb C}^{\ast })^g$  the module ${\hat M}_{\Lambda}$ is a free
${\cal T}_{g}$-module of rank $g$ generated by $g$ functions
from ${\hat M}_{\Lambda}(1)$.
\end{theorem}

Let $A$ be the ring of meromorphic functions on $\Gamma$
with poles only on the divisor $(f=0)$.

\begin{corollary}  For values of $h$ specified in Theorem \ref{th-2} there is an embedding of the ring
$$
 A \rightarrow Mat(g,{\cal T}_{g}).
$$
\end{corollary}

In the case $g = 2$ there is another way of identification of two lines on ${\mathbb C}P^1\times {\mathbb C}P^1$ which is suitable for our goals.

We set
\begin{equation} \label{eq10}
\Omega = {\mathbb C}P^1\times {\mathbb C}P^1 / \{([1,0],[t_1,t_2])\sim
([t_1,t_2],[0,1])\}.
\end{equation}
 Let $g,g_1,g_2$ are the following functions on ${\mathbb C}^4$:
$$
 g(z_1,z_2,w_1,w_2)=\alpha z_1w_1+\beta z_1w_2+\gamma z_2w_1+\delta
 z_2w_2, \; \alpha,\beta,\gamma,\delta\in{\mathbb C},
$$
$$
 g_i(z_1,z_2,w_1,w_2)=\alpha_i z_1w_1+\beta_i z_1w_2+\gamma_i
 z_2w_1+\delta_i z_2w_2,\ \alpha_i,\beta_i,\gamma_i,\delta_i\in{\mathbb C},
$$
such that
$$
g(0,1,0,1) \neq 0,
$$
\begin{equation} \label{eq11}
 g(1,0,t_1,t_2)-B g(t_1,t_2,0,1)=0,\;
\end{equation}
\begin{equation} \label{eq12}
 g_i(1,0,t_1,t_2)-c_ig_i(t_1,t_2,0,1)=0,\
 \end{equation}
where $ B,c_i\in {\mathbb C}^*$ are fixed constants.
Let us fix $\Lambda\in{\mathbb C}^*$.

The discrete Baker-Akhiezer module
${\hat M}_{\Omega,\Lambda}=\cup_{k=0}^\infty {\hat M}_{\Omega,\Lambda}(k)$ in this case is defined by
$$
 {\hat M}_{\Omega,\Lambda}(k)=\left\{\varphi=\frac{\tilde{h}(n_1,n_2,P)}{g(P)^k}\prod_{j=1}^2
 \left(\frac{g_j(P)}{g(P)}\right)^{n_j}\right\},
$$
where $\tilde{h}$ is a function of the form (\ref{eq9}) and $\varphi(n_1,n_2,P)$, $P\in {\mathbb C}^4,$ satisfies the identity
$$
 \varphi(n_1,n_2,1,0,t_1,t_2)-\Lambda\varphi(n_1,n_2,t_1,t_2,0,1)=0.
$$

\begin{theorem}\label{th-3}
For an uncountable number of $h\in ({\mathbb C}^{\ast })^2$ the module ${\hat M}_{\Omega,\Lambda}$ is a free ${\cal T}_2$-module of
 rank 2 generated by two functions from ${\hat M}_{\Omega,\Lambda}(1)$.
\end{theorem}

Let $A$ denote the ring of the meromorphic functions on $\Omega$
with poles only on the curve defined by the equation $g(P)=0$.

\begin{corollary}  For values of $h$ specified in Theorem \ref{th-3} there is a ring embedding
$$
 A\rightarrow Mat(2,{\cal T}_2).
$$
\end{corollary}

\section{Proofs}
Theorems \ref{th-1} to \ref{th-3} follow from their differential analogues. Since the schemes of the proofs are similar, 
we only  prove Theorem \ref{th-1} and Theorem \ref{th-2}.

\subsection{Proof of Theorem \ref{th-1}}
Let
\bea
&&
\nabla_i=\partial_i-\zeta_i(z),\quad \partial_i=\partial/\partial x_i.
\non
\ena
It is easy to see that it acts on $M_c$. Let ${\cal D}={\cal K}[\partial_1,...,\partial_g]$. Then $M_c$ is a ${\cal D}$-module.
It is called the Baker-Akhiezer module of $(X,\Theta)$ \cite{N1}. Let
\bea
&& {\rm gr}\,M_c=\oplus_i{\rm gr}_i M_c, \quad {\rm gr}_i M_c=M_c(i)/M_c(i-1).
\non
\ena
Since $\partial_i M_c(m)\subset M_c(m+1)$, ${\rm gr}M_c$ is also a
 ${\cal D}$-module. Recall that we assume that $\Theta$ is non-singular 
in this paper. Then the following theorem is proved in \cite{N1}.

\begin{theorem}
 The module ${\rm gr}M_c$ is a free ${\cal D}$-module of rank $g!$.
\end{theorem}

More precisely there exists a ${\cal D}$-free basis $\varphi_{ij}$ such that $\varphi_{ij}\in {\rm gr}_iM_c$, $1\leq i\leq g$, $1\leq j\leq r_j$ with
\bea
&&
r_i=i^g-(i-1)^g-\sum_{j=1}^{i-1}r_j\bc{g+i-j-1}{g-1}, \quad r\geq 2,
\non
\ena
and $r_1=1$.
Moreover, for each $i$, one can find $\varphi_{ij}$
in $\{F_{i,a}(z,x)\}$, that is, one can write
\bea
&&
\varphi_{ij}=F_{i,a_{ij}}(z,x)
\non
\ena
for some $a_{ij}\in {\mathbb Z}^g/i{\mathbb Z}^g$.

We remark that, in Theorem \ref{th-1}, $c=0$ is not excluded. 
This is because we consider ${\cal K}$, the space of meromorphic functions of $x$, as a coefficient field of ${\cal D}$ and $M_c$. 

Recall that $T_i$ acts also on $M_c$. It satisfies
\bea
&&
T_iM_c(m) \subset M_c(m+1).
\non
\ena
Therefore $T_i$ acts on ${\rm gr}M_c$ too. For $F(z,x)\in M_c$ we have the expansion
\bea
&&
T_i F(z,x)=F(z,x)+\nabla_i F(z,x) h_i+O(h_i^2),
\non
\ena
and it is possible to define the map ${\tilde T}_i=(T_i-1)/h_i:M_c\rightarrow
M_c$:
\bea
&&
{\tilde T}_i F(z,x)=\frac{1}{h_i}\left(T_i F(z,x)-F(z,x)\right).
\non
\ena
It satisfies
\bea
&&
{\tilde T}_i F(z,x)=\nabla_i F(z,x)+O(h_i).
\label{tti-expansion}
\ena
Notice that, as an action on ${\rm gr} M_c$,
\bea
&&
{\tilde T}_i=\frac{1}{h_i} T_i.
\non
\ena

We prove

\begin{theorem}\label{th-5} For an uncountable number of
$h\in ({\mathbb C}^\ast)^g$,
${\rm gr} M_c$ is a free ${\cal T}_g$-module of rank $g!$ with
a basis $\{\varphi_{ij}\}$.
\end{theorem}
\vskip2mm
\noindent
{\it Proof}.
Since ${\rm gr}M_c$ is a free ${\cal D}$ module, for each $k$,
the set of elements
\bea
&&
\partial_1^{k_1}\cdots\partial_g^{k_g}\varphi_{ij},
\non
\\
&&
k_1+\cdots+k_g=k'-i,
\quad
0\leq k'\leq k,
\quad
1\leq i\leq g,
\quad
1\leq j\leq r_i,
\label{der-phi-ij}
\ena
is a ${\cal K}$-basis of $M_c(k)$.
The number of elements (\ref{der-phi-ij}) is $N_k:=k^g$.
Let us enumerate them as $\psi^k_1$,...,$\psi^k_{N_k}$.

Expand
\bea
&&
\theta(z)^k \psi^k_i=\sum a^k_{i,{\bf \mu}}(x)z^{{\bf \mu}},
\quad
{\bf \mu}=(\mu_1,...,\mu_g).
\non
\ena
Since $\{\psi^k_i\}$ is linearly independent over ${\cal K}$,
there exist ${\bf \mu}^{(k,1)}$,...,${\bf \mu}^{(k,N_k)}$ such that
\bea
&&
\det\left(a^k_{i,{\bf \mu}^{(k,j)}}(x)\right)_{1\leq i,j\leq N_k}\neq 0,
\non
\ena
where "$\neq0$" signifies that it is not identically zero as a function
of $x$.

Consider correspondingly
\bea
&&
{\tilde T}_1^{k_1}\cdots{\tilde T}_g^{k_g}\varphi_{ij}.
\label{difference-phi-ij}
\ena
Let us denote the function in (\ref{difference-phi-ij}) which has the same
$(k_1,...,k_g)$ as $\psi^k_i$ by ${\tilde \psi}^k_i$.
Then
\bea
&&
{\tilde \psi}^k_i(z,x,h)=\psi^k_i(z,x)+\sum_{l=1}^g O(h_l)
\non
\ena
If we expand
\bea
&&
\theta(z)^k{\tilde \psi}^k_i=
\sum {\tilde a}^k_{i,{\bf \mu}}(x,h)z^{{\bf \mu}}.
\non
\ena
then
\bea
&&
{\tilde a}^k_{i,{\bf \mu}}(x,h)=a^k_{i,{\bf \mu}}(x)+\sum_{l=1}^g O(h_l),
\non
\ena
and
\bea
&&
\det\left({\tilde a}^k_{i,{\bf \mu}^{(k,j)}}(x,0)\right)=
\det\left(a^k_{i,{\bf \mu}^{(j)}}(x)\right)\neq 0,
\non
\ena
Notice that
$\det\left(a^k_{i,{\bf \mu}^{(k,j)}}(x)\right)$ is an analytic
function of $x$ and the zero set of it is of measure zero.
Thus
\bea
&&
{\mathbb C}^g\backslash\cup_{k=0}^\infty
\left(\det\left(a^k_{i,{\bf \mu}^{(k,j)}}(x)\right)=0\right)
\label{value-x}
\ena
has positive measure and contains an uncountable number of elements.
Take any $x_0$ from (\ref{value-x}).
Since ${\tilde a}^k_{i,{\bf \mu}}(x,h)$ is an analytic function of $(x,h)$, the set
\bea
&&
{\mathbb C}^g\backslash\cup_{k=0}^\infty
\{h\vert \det\left({\tilde a}^k_{i,{\bf \mu}^{(k,j)}}(x_0,h)\right)=0\}
\label{value-h}
\ena
contains an uncountable number of elements.  Moreover it contains
elements of the form $h_0=(h_{01},...,h_{0g})$, $h_{0i}\neq 0$ for any
$i$, since $\cup_{i=1}^g \{\sum_{j\neq i}h_je_j\in {\mathbb C}^g\vert h_j\in{\mathbb C}^g\}$
is also of measure zero. Take such $h_0$. Then, for any $k$,
\bea
&&
\det\left({\tilde a}^k_{i,{\bf \mu}^{(k,j)}}(x,h_0)\right)\neq 0.
\non
\ena
For such $h_0$ $\{{\tilde \psi}^{k}_i\}$ is linearly independent
 and generate $M_c(k)$ over ${\cal K}$ for all $k\geq 0$.
Therefore ${\rm gr}M_c$ is a free ${\cal T}_g$ module with
the basis $\{\varphi_{ij}\}$. $\Box$
\vskip2mm

Let us prove Theorem \ref{th-1}. Notice that $T_i$ satisfies
the following commutation relation with a function of $x$:
\bea
&&
T_i F(x)=F(x+h_ie_i) T_i.
\non
\ena
By definition the discretization ${\hat \varphi}_{ij}(n)$
of $\varphi_{ij}(z,x)$ is
\bea
&&
{\hat \varphi}(n)=T^n\varphi_{ij}(z,x).
\non
\ena
By Theorem \ref{th-5} any element of $M_c$ can uniquely be written as
a linear combinations of
\bea
&&
T^m\varphi_{ij}, \quad m\in {\mathbb Z}_{\geq 0}^g,
\quad
1\leq i\leq g,
\quad
1\leq j\leq r_i,
\non
\ena
with the coefficients in ${\cal K}$. The discretization of
 the element of the form
$f(x)T^m\varphi_{ij}$ with $f(x)\in {\cal K}$, is given by
\bea
&&
T^n\left(f(x)T^m\varphi_{ij}(z,x)\right)=f(x+nh)T^mT^n\varphi_{ij}(z,x)
={\hat f}(n)T^m {\hat \varphi}_{ij}(n).
\non
\ena
Thus any element of ${\hat M}_c$ can be written as
a linear combination of $\{T^m{\hat \varphi}_{ij}(n)\}$ with the
coefficients in ${\hat {\cal K}}$.

Moreover this description of an element
of ${\hat M}_c$ as a linear combination of
$\{T^m{\hat \varphi}_{ij}(n)\}$ is unique.
In fact, suppose that
\bea
&&
\sum {\hat f}_{ij}(n)T^{m_{ij}}{\hat \varphi}_{ij}(n)=0.
\label{discrete-lin-rel}
\ena
Applying $T^{-n}$ to (\ref{discrete-lin-rel}) we get
\bea
&&
\sum f_{ij}(x)T^{m_{ij}}\varphi_{ij}(z,x)=0.
\non
\ena
It follows that $f_{ij}(x)=0$  for any $(i,j)$, 
since $\{\varphi_{ij}(z,x)\}$
 is linearly independent over ${\cal K}$.
Consequently ${\hat f}_{ij}(n)=0$ for every $(i,j)$.

Thus ${\hat M}_c$ is proved to be a free ${\cal T}_g$ module
with a basis ${\hat \varphi}_{ij}$. $\Box$

\subsection{Proof of Theorem \ref{th-2}}
We shall firstly give a construction of functions $f$, $f_i$ satisfying
the conditions (\ref{eq2})-(\ref{eq6}) together with further conditions and
related functions ${\tilde f}_i$.

Consider the function $F(z_1,z_2,t,s)$, $t\in {\mathbb C}^g$,
$s\in {\mathbb C}$ of the form
\bea
&&
F(z_1,z_2,t,s)=\sum_{k=1}^g (\gamma_k(s)z_1+\delta_k(s)z_2)t_k,
\non
\ena
and the following equation for $F$:
\bea
&&
F(a_1,b_1,t,s)=Ae^sF(a_2,b_2,{\cal P}(t),s).
\label{Fs-eq}
\ena
This equation gives $g$ linear homogeneous equations for $2g$ unknown
variables $\gamma_k$, $\delta_k$. By examining the case $(a_1,b_1)=(1,0)$,
$(a_2,b_2)=(0,1)$, ${\cal P}(t)=(\lambda_1 t_1,...,\lambda_g t_g)$,
we see easily that, for generic choices of $a_i,b_i$, ${\cal P}$,
there exist $g$ linearly independent solutions
$\{F_i\}$ of (\ref{Fs-eq}) such that the following conditions are
satisfied.
\vskip2mm

\noindent
(i) $F_i(z_1,z_2,t,0)$ is independent of $i$. Set $f(z_1,z_2,t)=F_i(z_1,z_2,t,0)$.
\vskip2mm

\noindent
(ii) The set of functions $\{f, \partial_s F_i(z_1,z_2,t,0)\}$ is linearly independent.
\vskip2mm

\noindent
(iii) $f(a_1,b_1,v_j)\neq 0$ for any $j$.
\vskip2mm

We take ${\tilde c}_i, h_i\in {\mathbb C}^\ast$ and set
\bea
&&
c_i=Ae^{{\tilde c}_i h_i}.
\non
\ena

We define $f_i$ and ${\tilde f}_i$ by
\bea
f_i(z_1,z_2,t,h_i)&=&F_i(z_1,z_2,t,{\tilde c}_ih_i),
\label{fi}
\\
{\tilde f}_i(z_1,z_2,t)&=&{\tilde c}_i \partial_sF_i(z_1,z_2,t,0).
\label{fti}
\ena
Then $f$, $f_i$ satisfy (\ref{eq2})-(\ref{eq6}) and
${\tilde f}_i$ satisfies
\bea
&&
{\tilde f}_i(z_1,z_2,t)=\partial_{h_i}f_i(z_1,z_2,t,0),
\label{fti-2}
\\
&&
{\tilde f}_i(a_1,b_1,t)-A{\tilde f}_i(a_2,b_2,{\cal P}(t))
-{\tilde c}_i f(a_1,b_1,t)=0,
\label{eq-fti}
\ena
due to Equation (\ref{Fs-eq}).
Moreover, by the property (ii), $\{f, {\tilde f}_1,...,{\tilde f}_g\}$
is linearly independent.

Nextly we consider, for $k$ fixed, a function $h(x,P)$, $P\in {\mathbb C}^{g+2}$
such that
\bea
&&
h(x,P)=\sum_{0\leq j\leq k, |\alpha|=k}
h_{j\alpha}(x)z_1^jz_2^{k-j}t^\alpha,
\label{def-hx}
\\
&&
\frac{h(x,a_1,b_1,t)}{f(a_1,b_1,t)^k}-
\Lambda e^{-\sum_{i=1}^g {\tilde c}_i x_i}
\frac{h(x,a_2,b_2,{\cal P}(t))}{f(a_2,b_2,{\cal P}(t))^k}=0.
\label{eq-hx}
\ena

The equation (\ref{eq-hx}) is equivalent to the system of linear homogeneous
equations for $\{h_{j\alpha}\}$.
As shown in \cite{MM}
\bea
&&
\sharp\{h_{j\alpha}\}-\sharp\{\text{equations}\}=g\bc{g+k-1}{g},
\non
\ena
where $\sharp S$ denotes the number of elements of $S$.
Therefore the equation (\ref{eq-hx}) has non-trivial solutions.
Moreover it is possible take a basis of solutions such that each element of a basis is a rational function of $e^{\sum_{i=1}^g{\tilde c}_i x_i}$ and is analytic at $x=0$.

Let ${\cal K}$ be the ring of rational functions of
$e^{\sum_{i=1}^g{\tilde c}_i x_i}$ and
\bea
&&
M_\lambda=\cup_{k=0}^\infty M_{\lambda}(k),
\qquad
M_\Lambda(k)=
\left\{ \frac{h(x,P)}{f(P)^k} \right\},
\non
\ena
where $h(x,P)$ runs over functions which satisfy (\ref{def-hx}), (\ref{eq-hx})
and are rational functions of $e^{\sum_{i=1}^g{\tilde c}_i x_i}$. Obviously
$M_\Lambda$ is a vector space over ${\cal K}$.
As remarked above we can take a basis of each $M_\Lambda(k)$ over ${\cal K}$
such that each element of the basis is analytic at $x=0$.

Equation (\ref{eq-hx}) signifies that an element $\varphi(x,P)$ of $M_\Lambda$
satisfies
\bea
&&
\varphi(x,a_1,b_1,t)-\Lambda e^{-\sum_{i=1}^g{\tilde c}_i x_i}
\varphi(x,a_2,b_2,{\cal P}(t))=0.
\non
\ena

Let
\bea
&&
\xi_i(P)=\frac{{\tilde f}_i(P)}{f_i(P)}.
\non
\ena
Then it satisfies
\bea
&&
\xi_i(a_1,b_1,t)-\xi_i(a_2,b_2,{\cal P}(t))-{\tilde c}_i=0,
\label{eq-xii}
\ena
due to (\ref{eq-fti}).
We set
\bea
&&
\nabla_i=\partial_i+\xi_i(P).
\non
\ena
Using (\ref{eq-xii}) one can easily check that $\nabla_i$ acts on $M_\Lambda$
and satisfies $\nabla_iM_\Lambda(k)\subset M_{\Lambda}(k+1)$.
Thus $M_\Lambda$  and ${\rm gr}M_\Lambda$ become modules over
the ring of differential operators ${\cal D}:={\cal K}[\partial_1,...,\partial_g]$, where $\partial_i$ acts by $\nabla_i$. The ${\cal D}$ module
$M_\Lambda$ is the Baker-Akhiezer module of $(\Gamma, (f=0))$ constructed
in \cite{MM}.

The following theorem had been proved in \cite{MM}.

\begin{theorem}\label{BA-D-mod}
The module ${\rm gr}M_\Lambda$ is a free ${\cal D}$-module of rank $g$ generated
by $g$ functions from $M_\Lambda(1)$.
\end{theorem}

Similarly to the case of abelian varieties we define the operator
$T_i$ by
\bea
&&
T_i=\frac{f_i(P,h_i)}{f(P)}{\rm e}^{h_i\partial_i},
\non
\ena
where ${\rm e}^{h_i\partial_i}$ is the shift operator:
\bea
&&
{\rm e}^{h_i\partial_i}G(...,x_i,...)=G(...,x_i+h_i,...).
\non
\ena
By Equations (\ref{eq3}), (\ref{eq6}) $T_i$ acts on $M_\Lambda$ and satisfies
$T_iM_\Lambda(k)\subset M_\Lambda(k+1)$.
Therefore $T_i$ acts on ${\rm gr}M_\Lambda$.

By (\ref{fti-2}) we have
\bea
&&
f_i(P,h_i)=f(P)+{\tilde f}_i(P)h_i+O(h_i^2).
\non
\ena
Consequently
\bea
&&
T_i=1+h_i \nabla_i+O(h_i^2).
\non
\ena
We set
\bea
&&
{\tilde T}_i=\frac{1}{h_i}(T_i-1)=\nabla_i+O(h_i).
\non
\ena
On ${\rm gr}M_\Lambda$ we have
\bea
&&
{\tilde T}_i=\frac{1}{h_i}T_i.
\non
\ena

The discretization ${\hat M}_\Lambda$ of $M_\Lambda$ is similarly defined, using $T_i$, to the case of Abelian varieties.
Expcilitly ${\hat M}_\Lambda$ is given by (\ref{Mlh-gamma}) and (\ref{Mlhk-gamma}).

The proof of Theorem \ref{th-2} is completely similar to that of Theorem \ref{th-1} and reduces to Theorem \ref{BA-D-mod} using ${\tilde T}_i$. $\Box$

\section{Commuting difference operators}
In this section we give examples of explicit forms of 
commuting difference operators.

\subsection{Two-points operators: $g=1$}
Let $g=1$ in the Theorem \ref{th-1}, $X={\mathbb C}/({\mathbb Z}+\tau{\mathbb Z})$.
In this case the DBA-module ${\hat M}_0$ is generated over ${\cal T}_1$ by the function
$$
 \psi(n,z)=\frac{\theta(z+x+nh)}{\theta(z)}\left(\frac{\theta(z-h)}{\theta(z)}\right)^n\in {\hat M}_c(1).
$$
Let
$$
 \lambda=\frac{\theta(z-h)\theta(z+h)}{\theta^2(z)}.
$$
There is a unique operator of the form
$$
 L_1=v_2(n)T^2+v_1(n)T+v_0(n)
$$
such that
\begin{equation}\label{u4.1.1}
 L_1\psi(n,z)=\lambda(z)\psi(n,z).
\end{equation}
Let us find the coefficients $v_i(n)$.
We divide (\ref{u4.1.1}) by $\left(\theta(z-h)/\theta(z)\right)^n$ and
 multiply by $\theta(z)^3$:
\bea
&&
 v_2(n)\theta(z+x+(n+2)h)\theta^2(z-h)
+v_1(n)\theta(z+x+(n+1)h)\theta(z-h)\theta(z)
\non
\\
&&
 +v_0(n)\theta(z+x+nh)\theta(z)^2=
\theta(z-h)\theta(z+h)\theta(z+x+nh).
\label{u4.1.2}
\ena
We recall that $\theta(\frac{1}{2}+\frac{1}{2}\tau)=0$.
Let us substitute $z=p=\frac{1}{2}+\frac{1}{2}\tau+h$ in (\ref{u4.1.2}). 
We obtain $v_0=0$. 
Let us divide (\ref{u4.1.2}) by $\theta(z-h)$ and again substitute $z=p=\frac{1}{2}+\frac{1}{2}\tau+h$.
 We obtain

$$
 v_1(n)=\frac{\theta(p+x+nh)\theta(p+h)}{\theta(p+x+(n+1)h)\theta(p)}.
$$
We put $z=q=\frac{1}{2}+\frac{1}{2}\tau$ (\ref{u4.1.2}) and obtain
$$
 v_2(n)=\frac{\theta(q+x+nh)\theta(q+h)}{\theta(q+x+(n+2)h)\theta(q-h)}.
$$
Similarly, for
$$
 \mu=\frac{\theta(z-h)\theta^2(z+\frac{h}{2})}{\theta^3(z)}
$$
we have

\begin{equation}\label{u4.1.3}
 L_2\psi=\left(u_3(n)T^3+u_2(n)T^2+u_1(n)T+u_0(n)\right)\psi=\mu\psi.
\end{equation}
From (\ref{u4.1.3}) we obtain $u_0=0$, 

$$
 u_1(n)=\frac{\theta(p+x+nh)\theta^2(p+\frac{h}{2})}{\theta(p+x+(n+1)h)\theta^2(p)},\quad
 u_3(n)=\frac{\theta(q+x+nh)\theta^2(q+\frac{h}{2})}{\theta(q+x+(n+3)h)\theta^2(q-h)}.
$$
To find $u_2(n)$ let us substitute in (\ref{u4.1.3}) $z=r=\frac{1}{2}+\frac{1}{2}\tau-\frac{h}{2}$:

$$
 u_2(n)=-u_1(n)\frac{\theta(r+x+(n+1)h)\theta(r)}{\theta(r+x+(n+2)h)\theta(r-h)}-
 u_3(n)\frac{\theta(r+x+(n+3)h)\theta(r-h)}{\theta(r+x+(n+2)h)\theta(r)}.
$$
Operators $L_1$ and $L_2$ commute.

It is easy to see that for the meromorphic function $\eta$ with poles at $p$ and $q$ there
is an operator of the form
$$
  L=\sum^{N_+}_{i=N_-}v_i(n)T^i
$$
such that
$$
 L\psi=\eta\psi.
$$
We see that in the case $g=1$ our construction is involved in the two-points construction \cite{Kr}.

\subsection{$2\times 2$-matrix operators:  Abelian varieties}
Let $g=2$ in the Theorem \ref{th-1}, $X={\mathbb C}^2/({\mathbb Z}^2+\tau{\mathbb Z}^2)$.
The functions
\bea
 \psi_1&=&\frac{\theta(z+x+nh)}{\theta(z)}
\prod_{j=1}^2
 \left(\frac{\theta(z-h_je_j)}{\theta(z)}\right)^{n_j}\in {\hat M}_0(1),
\non
\\
 \psi_2&=&\frac{\theta(z+x+nh+\beta)\theta(z-\beta)}{\theta^2(z)}
\prod_{j=1}^2
 \left(\frac{\theta(z-h_je_j)}{\theta(z)}\right)^{n_j}\in {\hat M}_0(2)
\ena
gives a  basis in ${\hat M}_0$, where $\beta$ belongs to some open
everywhere dense subset in ${\mathbb C}^2$. Let us find the 
operator corresponding to the function

$$
 \lambda=\frac{\theta(z-h_1e_1)\theta(z+h_1e_1)}{\theta^2(z)}.
$$
We have

\begin{equation}\label{u4.2.1}
 L_{11}\psi_1+L_{12}\psi_2=\lambda\psi_1,
\end{equation}
$$
 L_{21}\psi_1+L_{22}\psi_2=\lambda\psi_2.
$$
Operators $L_{11}$ and $L_{12}$ have the form
$$
 L_{11}=v_{20}T_1^2+v_{11}T_1T_2+v_{02}T_2^2+v_1T_1+v_2T_2+v_0,\quad
 L_{12}=u_1T_1+u_2T_2+u_0.
$$
Let us divide (\ref{u4.2.1}) by
$
 \prod_{j=1}^2\left(\theta(z-h_je_j)/\theta(z)\right)^{n_j}
$  and multiply by $\theta(z)^3$. Then we get
\vskip2mm
\hskip10mm
$\displaystyle{
v_{20}\theta(z+x+nh+2h_1e_1)\theta(z-h_1e_1)^2
}$
\vskip2mm
\hskip8.5mm
$\displaystyle{
 +v_{11}\theta(z+x+(n+1)h)
\theta(z-h_1e_1)\theta(z-h_2e_2)
}$
\vskip2mm
\hskip8.5mm
$\displaystyle{
+v_{02}\theta(z+x+nh+2h_2e_2)\theta(z-h_2e_2)^2
}$
\vskip2mm
\hskip8.5mm
$\displaystyle{
+v_{1}\theta(z+x+nh+h_1e_1)\theta(z-h_1e_1)\theta(z)
}$
\vskip2mm
\hskip8.5mm
$\displaystyle{
+v_{2}\theta(z+x+nh+h_2e_2)\theta(z-h_2e_2)\theta(z)+
v_{0}\theta(z+x+nh)\theta(z)^2
}$
\vskip2mm
\hskip8.5mm
$\displaystyle{
+u_{1}\theta(z+x+nh+h_1e_1+\beta)\theta(z-\beta)\theta(z-h_1e_1)
}$
\vskip2mm
\hskip8.5mm
$\displaystyle{
+u_{2}\theta(z+x+nh+h_2e_2+\beta)\theta(z-\beta)
\theta(z-h_2e_2)
}$
\bea
&&
+u_{0}\theta(z+x+nh+\beta)\theta(z-\beta)\theta(z)
=\theta(z-h_1e_1)\theta(z+h_1e_1)\theta(z+x+nh).
\label{A1}
\ena

\begin{lemma}\label{lem-1}
The equalities
$$
v_0=u_0=0
$$
are valid.
\end{lemma}
{\it Proof.}
Let $p_1$ and $p_2$ be the points of intersection of the curves
$\theta(z-h_1e_1)=0$ and $\theta(z-h_2e_2)=0$.
Let us substitute $z=p_1$ and $z=p_2$ in (\ref{A1}):
$$
 v_{0}\theta(p_i+x+nh)\theta(p_i)^2+
 u_{0}\theta(p_i+x+nh+\beta)\theta(p_i-\beta)\theta(p_i)=0.
$$
These equations can be considered as a system of linear equations
for $v_0, u_0$.
If $v_0\ne 0$ or $u_0\ne 0$ then
\bea
 &&
\theta(p_1+x+nh)\theta(p_2+x+nh+\beta)\theta(p_2-\beta)\theta(p_1)-
\non
\\
&&
 \theta(p_2+x+nh)\theta(p_1+x+nh+\beta)\theta(p_1-\beta)\theta(p_2)=0.
\non
\ena
If $\beta$ is a solution of $\theta(p_1-\beta)=0$ then this equality is not valid. Consequently for $\beta$ in general position
this equality is not valid.  Thus Lemma \ref{lem-1} is proved. $\Box$
\vskip2mm

Let us restrict (\ref{A1}) on the curve $\theta(z-h_1e_1)=0$ and divide by $\theta(z-h_2e_2)$:

\begin{equation}\label{u4.2.3}
 v_{02}\theta(z+x+nh+2h_2e_2)\theta(z-h_2e_2)+v_{2}\theta(z+x+nh+h_2e_2)\theta(z)+
$$
$$
 u_{2}\theta(z+x+nh+h_2e_2+\beta)\theta(z-\beta)=0.
\end{equation}
Let $q_1$ and $q_2$ be the points of intersection of $\theta(z-h_1e_1)=0$ and $\theta(z)=0$. Then
$$
 v_{02}\theta(q_i+x+nh+2h_2e_2)\theta(q_i-h_2e_2)+u_{2}\theta(q_i+x+nh+h_2e_2+\beta)\theta(q_i-\beta)=0.
$$
By a similar argument as in the proof of Lemma \ref{lem-1} we obtain
$$
 v_{02}=v_2=u_2=0.
$$
We divide (\ref{A1}) by $\theta(z-h_1e_1)$ and get
$$
v_{20}\theta(z+x+nh+2h_1e_1)\theta(z-h_1e_1)+
v_{11}\theta(z+x+(n+1)h)\theta(z-h_2e_2)
$$
$$
+v_{1}\theta(z+x+nh+h_1e_1)\theta(z)
+u_{1}\theta(z+x+nh+h_1e_1+\beta)\theta(z-\beta)
$$
\begin{equation}\label{u4.2.5}
= \theta(z+x+nh)\theta(z+h_1e_1).
\end{equation}
Let us substitute $z=p_1$ and $z=p_2$ in (\ref{u4.2.5}). Then we obtain
$$\left(
  \begin{array}{c}
v_1\\
u_1\\
\end{array}\right)=A_1^{-1}
\left(
  \begin{array}{c}
\theta(p_1+x+nh)\theta(p_1+h_1e_1)\\
\theta(p_2+x+nh)\theta(p_2+h_1e_1)\\
\end{array}\right),
$$
$$
A_1=\left(
  \begin{array}{cc}
\theta(p_1+x+nh+h_1e_1)\theta(p_1)&\theta(p_1+x+nh+h_1e_1+\beta)\theta(p_1-\beta)\\
\theta(p_2+x+nh+h_1e_1)\theta(p_2)&\theta(p_2+x+nh+h_1e_1+\beta)\theta(p_2-\beta)
\end{array}\right).
$$
Let $r_1$ and $r_2$ be the points of intersection of $\theta(z)=0$ and $\theta(z-\beta)=0$. From (\ref{u4.2.5}) we obtain
$$\left(
  \begin{array}{c}
v_{20}\\
v_{11}\\
\end{array}\right)=A_2^{-1}
\left(
  \begin{array}{c}
\theta(r_1+x+nh)\theta(r_1+h_1e_1)\\
\theta(r_2+x+nh)\theta(r_2+h_1e_1)\\
\end{array}\right),
$$
$$
 A_2 =
 \left(
  \begin{array}{cc}
\theta(r_1+x+nh+2h_1e_1)\theta(p_1-h_1e_1)&
\theta(r_1+x+(n+1)h)\theta(r_1-h_2e_2)\\
\theta(r_2+x+nh+2h_1e_1)\theta(p_2-h_1e_1)&
\theta(r_2+x+(n+1)h)\theta(r_2-h_2e_2)
\end{array}\right).
$$

Similarly it is possible to find operators $L_{21}$, $L_{22}$ and an operator corresponding to
$$
 \frac{\theta(z-h_2e_2)\theta(z+h_2e_2)}{\theta^2(z)}.
$$

\subsection{$2\times 2$-matrix operators with rational coefficients}
It is well known fact that the Lame identity
$$
 (\partial_x^2-2\wp(x))\psi(x,z)=\wp(z)\psi(x,z),
$$
$$
 \psi(x,z)=\frac{\sigma(z+x)}{\sigma(x)\sigma(z)}e^{-x\zeta(z)},
$$
where $\sigma,\zeta,\wp$ are Weierstrass functions of the elliptic curve
$w^2=4y^3+\alpha_1y+\alpha_0$, becomes the form
$$
 \left(\partial_x^2-\frac{2}{x^2}\right)\psi^{\vee}(x,z)=\frac{1}{z^2}\psi^{\vee}(x,z),
$$
$$
 \psi^{\vee}(x,z)=\frac{z+x}{xz}e^{-\frac{x}{z}}
$$
under the degeneration $\alpha_i\rightarrow 0$.
The Lame potential becomes rational function $-\frac{2}{x^2}$. In this section we shall consider spectral variety $X^{\vee}$ obtained from the
Abelian variety
$X={\mathbb C}^2/({\mathbb Z}^2+\tau{\mathbb Z}^2)$ by a similar degeneration. Elements of the
corresponding DBA-module are expressed in terms of elementary functions,
coefficients of commuting difference operators are rational functions.
To describe $X^{\vee}$ we
recall Mumford's construction of the affine part of the Jacobian variety
of a hyperelliptic curve $\Sigma$ of genus $g$ (see \cite{N3}):
$$
 w^2=f(y)=4y^{2g+1}+\alpha_{2g}y^{2g}+\dots+\alpha_0.
$$
Let us introduce polynomials
$$
 a(y)=\sum_{i=1}^ga_{2i+1}y^{g-i},\ b(y)=\sum_{i=0}^gb_{2i}y^{g-i},\ c(y)=\sum_{i=0}^{g+1}c_{2i}y^{g+1-i},
$$
$b_0=1,c_0=4,a_1=0.$ We shall consider the affine space
${\mathbb C}^{3g+1}$ with the coordinates $(a_{2i+1},b_{2i},c_{2i})$.
The affine part $J(\Sigma)\backslash \Theta$ is given in ${\mathbb C}^{3g+1}$ by the following system of equations for $a_{2i+1},b_{2i},c_{2i}$:
$$
 a^2(y)+b(y)c(y)=f(y).
$$
In the case $g=2$ we have the following equations
$$
 \alpha_0-a_5^2-b_4c_6=0,\ \alpha_1-2a_3a_5-b_4c_4-b_2c_6=0,
$$
$$
 \alpha_2-a_3^2-b_4c_2-b_2c_4-c_6=0,\ \alpha_3-4b_4-b_2c_2-c_4=0,
\  \alpha_4-4b_2-c_2=0.
$$
We define spectral variety $X^{\vee}$ by the conditions $\alpha_i=0$.
From the last three equations one can find $c_2,c_4,c_6,$
and substitute it in first two equations.
One get that $X^{\vee}$ is isomorphic to the variety given in ${\mathbb C}^4$ by two equations
\begin{equation}\label{degenerate-eq1}
 b_4(a_3^2+4b_2^3-8b_2b_4)-a_5^2=0,
\quad
 a_3^2b_2-2a_3a_5+4(b_2^4-3b_2^2b_4+b_4^2)=0.
\end{equation}

Analytically this degeneration of the Jacobian variety is well described
by using the sigma function of $X$.
The sigma function is a certain modification of the Riemann's theta function which is originally introduced by Klein \cite{K1,K2}.
 The important
property for us now is that the sigma function $\sigma(z_1,z_2)$ becomes
 the Schur function
$$
 \sigma^{\vee}=\frac{z_1^3}{3}-z_2
$$
under the limit $\alpha_i\rightarrow 0$ \cite{BEL1,N4}. The $a_i,b_i,c_i$
coordinates of the Jacobian can explicitly be described using the sigma function (see \cite{N3}) and, consequently,
those of the variety given by (\ref{degenerate-eq1}) is described by the Schur function.

One can replace the Riemann's theta function by the sigma function in the
description of the DBA-module on $X$ in the previous section.
A free basis of the DBA-module is given by
\bea
 \psi_1&=&\frac{\sigma(z+x+nh)}{\sigma(z)}
\prod_{j=1}^2
 \left(\frac{\sigma(z-h_je_j)}{\sigma(z)}\right)^{n_j}\in {\hat M}_0(1),
\non
\\
 \psi_2&=&\frac{\sigma(z+x+nh+\beta)\sigma(z-\beta)}{\sigma^2(z)}
\prod_{j=1}^2
 \left(\frac{\sigma(z-h_je_j)}{\sigma(z)}\right)^{n_j}\in {\hat M}_0(2).
\non
\ena

Taking the limit $\alpha_i\rightarrow 0$ we get a new free DBA-module on
$X^{\vee}$ generated by the functions
\bea
 \psi_1^{\vee}&=&\frac{\sigma^{\vee}(z+x+nh)}{\sigma^{\vee}(z)}
\prod_{j=1}^2
 \left(\frac{\sigma^{\vee}(z-h_je_j)}{\sigma^{\vee}(z)}\right)^{n_j},
\non
\\
 \psi_2^{\vee}&=&\frac{\sigma^{\vee}(z+x+nh+\beta)\sigma^{\vee}(z-\beta)}{(\sigma^{\vee}(z))^2}
\prod_{j=1}^2
 \left(\frac{\sigma^{\vee}(z-h_je_j)}{\sigma^{\vee}(z)}\right)^{n_j}.
\non
\ena
Let $\Psi={}^t(\psi_1,\psi_2)$ and $\Psi^{\vee}=(\psi_1^{\vee}, \psi_2^{\vee})$.
As a limit of the identity $L(\lambda)\Psi=\lambda\Psi$ we get  $L^{\vee}(\lambda^{\vee})\Psi^{\vee}=\lambda^{\vee}\Psi^{\vee},$ where $\lambda^{\vee}$ is the corresponding limit of $\lambda$. For different  $\lambda^{\vee}$ and  $\mu^{\vee}$
operators $L^{\vee}(\lambda^{\vee})$ and  $L^{\vee}(\mu^{\vee})$ commute. By the method explained in the previous section we can directly compute
the operator corresponding to the function
$$
 \lambda^{\vee}=\frac{\sigma^{\vee}(z-h_1e_1)\sigma^{\vee}(z+h_1e_1)}{(\sigma^{\vee}(z))^2}=
 \frac{((z_1-h_1)^3/3-z_2)((z_1+h_1)^3/3-z_2)}{(z_1^3/3-z_2)^2}.
$$
For simplicity we put $h_1=h_2=1, x=0,\beta=(1,1/3)$. We have
$$
 L^{\vee}_{11}(\lambda^{\vee})=v_{20}T_1^2+v_{11}T_1T_2+v_1T_1,\quad L^{\vee}_{12}(\lambda^{\vee})=u_1T_1,
$$

$\displaystyle{
 u_1=\frac{-2n_1^2(n_1+1)(n_1+2)(n_1(n_1+3)+5)+6n_2(2n_1(n_1+1)(n_1+2)-3)-18n_2^2}{(n_1+2)(6n_2+n_1(n_1(n_1+6)+13)+14)},
}$
\vskip2mm
$\displaystyle{
 v_{20}=-u_1-\frac{n_1}{n_1+2},
\quad v_{11}=\frac{(n_1+2)(2n_1(n_1+1)-u_1(n_1+3))-6n_2}{3(n_1+2)(n_1+1)},
}$
\vskip2mm
$\displaystyle{
 v_1=2-v_{11}-\frac{2}{n_1+2},
}$
\bea
 L^{\vee}_{21}(\lambda^{\vee})&=&q_{30}T_1^3+q_{21}T_1^2T_2
+q_{12}T_1T_2^2+q_{20}T_1^2+q_{11}T_1T_2+q_1T_1,
\non
\\
 L^{\vee}_{22}(\lambda^{\vee})&=&p_{20}T_1^2+p_{11}T_1T_2+p_1T_1,
\non
\ena

$\displaystyle{
  p_1=\frac{2(n_1^6+9n_1^5+37n_1^4+48+n_1^2(106-27n_2))}{3(n_1+2)(n_1^3+6n_1^2+13n_1+6n_2+14)}
}$\par
\hskip10mm
$\displaystyle{
 +\frac{2(n_1(88-21n_2)+n_1^3(83-6n_2)+21n_2+9n_2^2)}{3(n_1+2)(n_1^3+6n_1^2+13n_1+6n_2+14)},
}$
\vskip2mm
$\displaystyle{
  p_{11}=\frac{2(5+n_1(11+n_1(n_1+6))-3n_2)-9(n_1+1)(n_1+2)q_{12}}{3(n_1+2)(n_1+3)},
}$
\vskip2mm
$\displaystyle{
 q_1=\frac{3p_{12}+3p_1-4+n_1(p_{11}+p_1-2)}{3(n_1+1)}+q_{12},
}$
\vskip2mm
$\displaystyle{
 p_{20}=\frac{-2n_1^6-24n_1^5-123n_1^4+12n_1^3(n_2-28)}{(n_1+3)(n_1^3+9n_1^2+28n_1+6n_2+34)}+
}$
\vskip1mm
\hskip10mm
$\displaystyle{
 +\frac{n_1^2(72n_2-501)+6n_1(21n_2-64)-18(n_2^2-2n_2+7)}{(n_1+3)(n_1^3+9n_1^2+28n_1+6n_2+34)},
}$
\vskip2mm
$\displaystyle{
 q_{11}=-q_1-q_{12},
\quad q_{20}=\frac{3(n_1+1)(q_{12}-q_1)}{n_1+3}-q_{21},
\quad q_{30}=\frac{2}{n_1+3}-p_{20}-1,
}$
\vskip2mm
$\displaystyle{
 q_{21}=\frac{9(n_1(n_1(n_1+3)+3)-3n_2-5)q_{12}+((n_1+3)^3-3n_2)q_{30}}{3(5+n_1(n_1(n_1+6)+12)-3n_2)},
}$
\vskip2mm
$\displaystyle{
 q_{12}=\frac{2(46+61n_1^4+12n_1^5+n_1^6+n_1(161-66n_2))}{9(n_1+2)(n_1^3+6n_1^2+13n_1+20)}+
}$
\vskip1mm
\hskip10mm
$\displaystyle{
 +\frac{2(n_1^3(163-6n_2)-21n_2+9n_2^2-4n_1^2(9n_2-58))}{9(n_1+2)(n_1^3+6n_1^2+13n_1+20)}.
}$
\vskip2mm
Similarly, for the function
$$
 \mu^{\vee}=\frac{\sigma^{\vee}(z-h_2e_2)\sigma^{\vee}(z+h_2e_2)}{(\sigma^{\vee}(z))^2}=
 \frac{(z_1^3/3-(z_2-h_2))(z_1^3/3-(z_2+h_2))}{(z_1^3/3-z_2)^2}
$$
we have
$$
 L_{11}^{\vee}(\mu^{\vee})=f_{11}T_1T_2+f_{02}T_2^2+f_2T_2,
\quad  L_{12}^{\vee}(\mu^{\vee})=g_2T_2,
$$

$\displaystyle{
 f_{11}=\frac{18n_1}{n_1(n_1(n_1+3)+4)+6(n_2+2)},\qquad f_{02}=\frac{f_{11}(n_1+2)}{3n_1}-1,
}$
\vskip2mm
$\displaystyle{
f_2=1-f_{02},\qquad g_2=-f_{11},
}$
\bea
 L_{21}^{\vee}(\mu^{\vee})&=&r_{21}T_1^2T_2+r_{12}T_1T_2^2+r_{03}T_2^3+r_{11}T_1T_2+r_{02}T_2^2+r_2T_2,
\non
\\
 L_{22}^{\vee}(\mu^{\vee})&=&j_{11}T_1T_2+j_{02}T_2^2+j_2T_2,
\non
\ena

$\displaystyle{
 r_{21}=\frac{18(n_1+1)}{n_1^3+6n_1^2+13n_1+20+6n_2},
\quad
 r_{03}=\frac{2(n_1+2)}{n_1^3+3n_1^2+4n_1+6(n_2+2)},
}$
\vskip2mm
$\displaystyle{
 r_{12}=\frac{9n_1r_{03}+9n_1^2r_{03}+6r_{21}+5n_1r_{21}+n_1^2r_{21}}{3n_1^2+9n_1+6},
}$
\vskip2mm
$\displaystyle{
 r_{11}=\frac{-n_1^3r_{12}-6(n_2+2)r_{12}+n_1^2(9r_{03}-6r_{12}+r_{21})+n_1(3r_{21}-9r_{03}-7r_{12})}{n_1^3+3n_1^2+4n_1+6(n_2+2)},
}$
\vskip2mm
$\displaystyle{
 r_{02}=-\frac{2(n_1^3+3n_1^2+4n_1+15+6n_2)r_{03}}{n_1^3+3n_1^2+4n_1+6(n_2+2)},
\quad r_2=-r_{02}-r_{03},
\quad j_{11}=-r_{21},
}$
\vskip2mm
$\displaystyle{
 j_{02}=-\frac{2+n_1+3n_1r_{03}}{n_1+2},
\quad j_2=\frac{n_1+2-j_{02}(n_1+2)-3n_1r_{02}-6n_1r_{03}}{n_1+2}.
}$
\vskip5mm

\subsection{$2\times 2$-matrix operators: rational spectral variety}
Let us consider the DBA-module ${\hat M}_{\Omega,1}$ of the case of $\Lambda=1$. 
We set
$$
 g=z_1w_1+z_1w_2+z_2w_2,
\ g_1=4z_1w_1+2z_1w_2+z_2w_2,
\ g_2=z_1w_1-z_1w_2+z_2w_2.
$$
Here $B=1$ in the formula (\ref{eq11}), and $c_1=2,c_2=-1$ in the formula (\ref{eq12}).
We choose the following basis of ${\hat M}_{\Omega,1}$:
\bea
 \psi_1&=&\frac{z_2w_1}{g}\left(\frac{g_1}{g}\right)^{n_1}\left(\frac{g_2}{g}\right)^{n_2},
\non
\\
 \psi_2&=&\frac{z_1w_1+(-1)^{n_2}2^{n_1}z_1w_2+2^{2n_1}z_2w_2}{g}\left(\frac{g_1}{g}\right)^{n_1}\left(\frac{g_2}{g}\right)^{n_2}.
\non
\ena
We have
$$
 \psi_i(n_1,n_2,[1,0],[t_1,t_2])- \psi_i(n_1,n_2,[t_1,t_2],[0,1])=0.
$$
Let
$$
 \lambda_1=\frac{z_2w_1}{g},\ \lambda_2=\frac{z_1z_2w_1w_2}{g^2}.
$$
These functions satisfy the identity
$$
 \lambda_i([1,0],[t_1,t_2])-\lambda_i([t_1,t_2],[0,1])=0.
$$
It is easy to check that
$$
\left(
  \begin{array}{cc}
 T_1+a_2T_2+a & b_1T_1+b_2T_2+b\\
 c_2T_2+c & d_1T_1+d_2T_2+d\\
\end{array}\right)
\left(
  \begin{array}{c}
\psi_1\\
\psi_2\\
\end{array}\right)=\lambda_1\left(
\begin{array}{c}
\psi_1\\
\psi_2\\
\end{array}\right),
$$
where
\vskip2mm
$\displaystyle{
  a_2=-1+(-2+(-1)^{n_2}2^{n_1}+3\cdot 2^{1+2n_1})b_1,\quad a=-4-a_2-b_1,
}$
\vskip2mm
$\displaystyle{
 b_1=\frac{3}{-1+4^{1+n_1}},\quad  b_2=\frac{3(-1+(-1)^{n_2}2^{1+n_1})}{(1+(-1)^{n_2}2^{n_1})(-1+4^{1+n_1})},\quad \ b=-4b_1-b_2,
}$
\vskip2mm
$\displaystyle{
  d_1=\frac{-1+4^{n_1}}{-1+4^{1+n_1}},\quad c_2=\frac{1}{2}(1-(-1)^{n_2}2^{n_1})+(-2+(-1)^{n_2}2^{n_1}+3\cdot2^{1+2n_1})d_1,
}$
\vskip2mm
$\displaystyle{
 c=1-c_2-d_1, \quad
 d_2=\frac{(-1+(-1)^{n_2}2^{1+n_1})d_1}{1+(-1)^{n_2}2^{n_1}},
\quad d=-4d_1-d_2.
}$
\vskip2mm

In a similar way we get
$$
\left(
  \begin{array}{cc}
L_{11} & L_{12}\\
L_{21} & L_{22}\\
\end{array}\right)
\left(
  \begin{array}{c}
\psi_1\\
\psi_2\\
\end{array}\right)=\lambda_2\left(
\begin{array}{c}
\psi_1\\
\psi_2\\
\end{array}\right),
$$
\vskip2mm
$\displaystyle{
L_{11}=-\frac{1}{2}T_1T_2+a'_{22}T_2^2+\frac{1}{2}T_1+a'_2T_2+a',
}$
\vskip2mm
$\displaystyle{
L_{12}=b'_{12}T_1T_2+b'_{22}T_2^2+b'_1T_1+b'_2T_2+b',
}$
\vskip2mm
$\displaystyle{
L_{21}=c'_{22}T_2^2+c'_2T_2+c',\quad
L_{22}=d'_{12}T_1T_2+d'_{22}T_2^2+d'_1T_1+T_2+d',
}$
\vskip2mm
where
\vskip2mm
$\displaystyle{
 a'_{22}=\frac{1}{2}(1-4(1+(-1)^{n_2}2^{1+3n_1}-3\cdot4^{n_1})b'_{12}+(-1)^{n_2}2^{n_1}b'_2-(-1)^{n_2}8^{n_1}b'_2),
}$
\vskip2mm
$\displaystyle{
 a'_2=(2-2a'_{22}-b'_{12}-2(-1)^{n_2}2^{n_1}b'_{12}),\quad a'=-a'_2-a'_{22},
\quad b'_{12}=\frac{3}{2(-1+4^{1+n_1})},
}$
\vskip2mm
$\displaystyle{
 b'_{22}=\frac{1}{2}(-4(1+(-1)^{n_2}2^{1+n_1})b'_{12}-(1+(-1)^{n_2}2^{n_1})b'_2),
\quad b'_1=-b'_{12},\
}$
\vskip2mm
$\displaystyle{
 b'_2=\frac{3}{-1+4^{n_1}},\quad b'=-b'_2-b'_{22},
}$
\vskip2mm
$\displaystyle{
 c'_{22}=-\frac{1}{4}(-1+(-1)^{n_2}2^{n_1})(-1+(-8-(-1)^{n_2}2^{3+n_1}+4^{2+n_1})d'_{12}
}$\par
\hskip10mm
$\displaystyle{
 +(-1)^{n_2}2^{1+n_1}(1+(-1)^{n_2}2^{n_1})),
}$
\vskip2mm
$\displaystyle{
 c'_{2}=-\frac{1}{2}-2c'_{22}-(1+(-1)^{n_2}2^{1+n_1})d'_{12},
\quad c'=-c'_2-c'_{22},
\quad d'_1=-d'_{12},
}$
\vskip2mm
$\displaystyle{
 d'_{12}=\frac{-1+4^{n_1}}{2(-1+4^{1+n_1})},\quad d'=-1-d'_{22},
}$
\vskip2mm
$\displaystyle{
 d'_{22}=\frac{1}{2}(-4(1+(-1)^{n_2}2^{1+n_1})d'_{12}-(1+(-1)^{n_2}2^{n_1})).
}$
\vskip5mm
\noindent {\bf Acknowledgments}

\noindent This research is partially supported by JSPS Grant-in-Aid for Scientific Research (C) 23540245.
The first author is also partially supported by RFBR grant 11-01-12106-ofi-m-2011. The second author would like to thank Koji Cho for useful suggestions on the proofs of theorems.
A part of this work was done while the the first author stayed in Tsuda College. He is grateful to this institution for kind hospitality.

\end{document}